\documentclass[prx,twocolumn,superscriptaddress]{revtex4-1}
\usepackage{}
\usepackage{amssymb}
\usepackage{amsfonts}
\usepackage{graphics,graphicx,epsfig,bm,amsmath,amsthm,amssymb}
\usepackage{bm}
\usepackage{bbm}
\usepackage{longtable}
\usepackage{multirow}
\usepackage{array}
\usepackage{color}
\usepackage[usenames,dvipsnames]{xcolor}
\usepackage{amsthm}
\usepackage{appendix}
\usepackage{mathrsfs}

\usepackage{float}
\usepackage[colorlinks=true,
linkcolor=blue,citecolor=blue,
urlcolor={blue},
pdfauthor={ },
pdftitle={ },
pdfsubject={ },
pdfkeywords={ }]{hyperref}

\makeatletter
\renewcommand{\maketag@@@}[1]{\hbox{\m@th\normalsize\normalfont#1}}%
\makeatother

\begin{document}

\title{Searching spin-mass interaction using a diamagnetic levitated magnetic resonance force sensor }

\author{Fang Xiong}
\affiliation{National Laboratory of Solid State Microstructures and Department of Physics, Nanjing University, Nanjing 210093, China}

\author{Tong Wu}
\affiliation{National Laboratory of Solid State Microstructures and Department of Physics, Nanjing University, Nanjing 210093, China}

\author{Yingchun Leng}
\affiliation{National Laboratory of Solid State Microstructures and Department of Physics, Nanjing University, Nanjing 210093, China}

\author{Rui Li}
\affiliation{Hefei National Laboratory for Physical Sciences at the Microscale and Department of Modern Physics, University of
Science and Technology of China, Hefei 230026,
China}
\affiliation{CAS Key Laboratory of Microscale Magnetic Resonance, University of Science and Technology of China, Hefei 230026,
China}
\affiliation{Synergetic Innovation Center of Quantum Information and Quantum Physics, University of Science and Technology of
China, Hefei 230026, China}

\author{Chang-Kui Duan}
\affiliation{Hefei National Laboratory for Physical Sciences at the Microscale and Department of Modern Physics, University of
Science and Technology of China, Hefei 230026,
China}
\affiliation{CAS Key Laboratory of Microscale Magnetic Resonance, University of Science and Technology of China, Hefei 230026,
China}
\affiliation{Synergetic Innovation Center of Quantum Information and Quantum Physics, University of Science and Technology of
China, Hefei 230026, China}

\author{\\Xi Kong}
\email{kongxi@nju.edu.cn}
\affiliation{National Laboratory of Solid State Microstructures and Department of Physics, Nanjing University, Nanjing 210093, China}

\author{Pu Huang}
\email{hp@nju.edu.cn}
\affiliation{National Laboratory of Solid State Microstructures and Department of Physics, Nanjing University, Nanjing 210093, China}

\author{Zhengwei Li}
\affiliation{Key Laboratory of Particle Astrophysics, Institute of High Energy Physics, Chinese Academy of Sciences, Beijing 100049, China}

\author{Yu Gao}
\affiliation{Key Laboratory of Particle Astrophysics, Institute of High Energy Physics, Chinese Academy of Sciences, Beijing 100049, China}

\author{Xing Rong}
\affiliation{Hefei National Laboratory for Physical Sciences at the Microscale and Department of Modern Physics, University of
Science and Technology of China, Hefei 230026,
China}
\affiliation{CAS Key Laboratory of Microscale Magnetic Resonance, University of Science and Technology of China, Hefei 230026,
China}
\affiliation{Synergetic Innovation Center of Quantum Information and Quantum Physics, University of Science and Technology of
China, Hefei 230026, China}

\author{Jiangfeng Du}
\email{djf@ustc.edu.cn}
\affiliation{Hefei National Laboratory for Physical Sciences at the Microscale and Department of Modern Physics, University of
Science and Technology of China, Hefei 230026,
China}
\affiliation{CAS Key Laboratory of Microscale Magnetic Resonance, University of Science and Technology of China, Hefei 230026,
China}
\affiliation{Synergetic Innovation Center of Quantum Information and Quantum Physics, University of Science and Technology of
China, Hefei 230026, China}

\begin{abstract}
Axion-like particles (ALPs) are predicted to mediate exotic interactions between spin and mass. We propose an ALP-searching experiment based on the levitated micromechanical oscillator, which is one of the most sensitive sensors for spin-mass forces at a short distance.  The proposed experiment tests the spin-mass resonant interaction between the polarized electron spins and a diamagnetically levitated microsphere. By periodically flipping the electron spins, the contamination from nonresonant background forces can be eliminated. The levitated microoscillator can  prospectively enhance the sensitivity by nearly $10^3$ times over current experiments for ALPs with mass in the range 4 meV to 0.4 eV.
\end{abstract}
\maketitle

\section{INTRODUCTION}
Light pseudoscalars exist in a number of beyond the Standard Model theories. One well motivated example is the axion \cite{Weinberg_PhysRevLett_1978, Wilczek_PhysRevLett_1978}, which is introduced via spontaneously broken the Peccei-Quinn (PQ) $U(1)$ symmetry  \cite{Peccei_PhysRevLett_1977, Peccei_PhysRevD_1977} to solve the strong CP problem, and is also a low-mass candidate for the dark matter in the universe \cite{Asztalos_AnnurevNucl_2006}. Generalized axion-like particles (ALPs) rise from dimensional compactification in string theory, which share similar interaction with electromagnetic fields and share a similar phenomenological role with the axions \cite{Arvanitaki_PhysRevD_2010,Cicoli_JHEP_2012, Anselm_PhysLettB_1982}. Motivated by axion and ALP's potential role in particle physics and cosmology, a number of experimental methods and techniques have been developed over the past few decades, such as the method proposed by  Moody and Wilczek  to detect cosmic axion \cite{Krauss_1985_PhysRevLett}, the photon-axion-photon conversion light shining through wall experiments  \cite{Redondo_ContemporaryPhysics_2011,Ballou_PhysRevD_2015}, the axion emission from the Sun \cite{Sikivie_PhysRevLett_1983,CAST_NatPhys_2017}, the dichroism and birefringence effects in external fields\cite{Raffelt_PhysRevD_1988, Fouche_PhysRevD_2016}, and the light pseudoscalar mediated macroscopic mass-mass\cite{Adelberger_PhysRevLett_2007}, spin-mass \cite{Wineland_PhysRevLett_1991,Venema_PhysRevLett_1992,Youdin_physrevlett_1996,Ni_PhyRrevLett_1999,Tullney_PhysRevLett_2013,Bulatowicz_PhysRevLett_2013, Crescini_PhysLetbB_2017,Rong_NatCommun_2018} and spin-spin  \cite{Vasilakis_PhysRevLett_2009,Terrano_PhysRevLett_2015} forces.

The pseudoscalar exchange between fermions results in spin-dependent forces \cite{Moody_PhysRevD_1984}. Most prior works detecting exotic spin-dependent forces \cite{Youdin_physrevlett_1996,Ni_PhyRrevLett_1999,Hammond_PhysRevLett_2007,Tullney_PhysRevLett_2013,Bulatowicz_PhysRevLett_2013,Arvanitaki_PhysRevLett_2014} are focused on the so-called axion window \cite{Turner_PhysicsReports_1990}, where the interaction range is  $200\mu$m--$20$cm. A (pseudo)scalar obtains a nonzero mass from the minima in its potential. In case of the axion, instanton-induced potential breaking its shift symmetry, which resolves the strong CP problem and explains the Universe's dark matter via misalignment mechanism \cite{Preskill_1983,Abbott_1983,Dine_1983}. It is desirable to find experimental techniques to search for such anomalous spin-dependent interactions at even shorter distances\cite{Ledbetter_PhysRevLett_2013}.

The levitated micromechanical and nanomechanical oscillators have been demonstrated as one of the ultrasensitive force sensors \cite{Gieseler_NP_2013,Ranjit_PRA_2016,Slezak_NJP_2018,Ahn_PRL_2018,Ricci_NL_2019,Monteiro_PRA_2020,David_2021} due to its ultralow dissipation and small size. It is one of the ideal methods to measure short-range force \cite{Geraci_PhysRevLett_2010,Chang_PNAS_2010,Ether_EPL_2015,Rider_PhysRevLett_2016,Hebestreit_PRL_2018,Winstone_PhysRevA_2018} with high precision. However, in short-ranged force measurements, surface noises from the electric static force fluctuation, the Casimir force and magnetic force limit the final sensitivity.

Here we propose a new method to investigate the spin-mass interaction using an ensemble of electron spins and a levitated diamagnetic microsphere mechanical oscillator. By periodically flip of the electron spins at the resonant frequency with the mechanical oscillator, the postulated force between electron spins and the microsphere mass is preserved while the spin-independent force noise from the surface is eliminated.

\section{SCHEME}
\begin{figure}
\includegraphics[width=7.4cm]{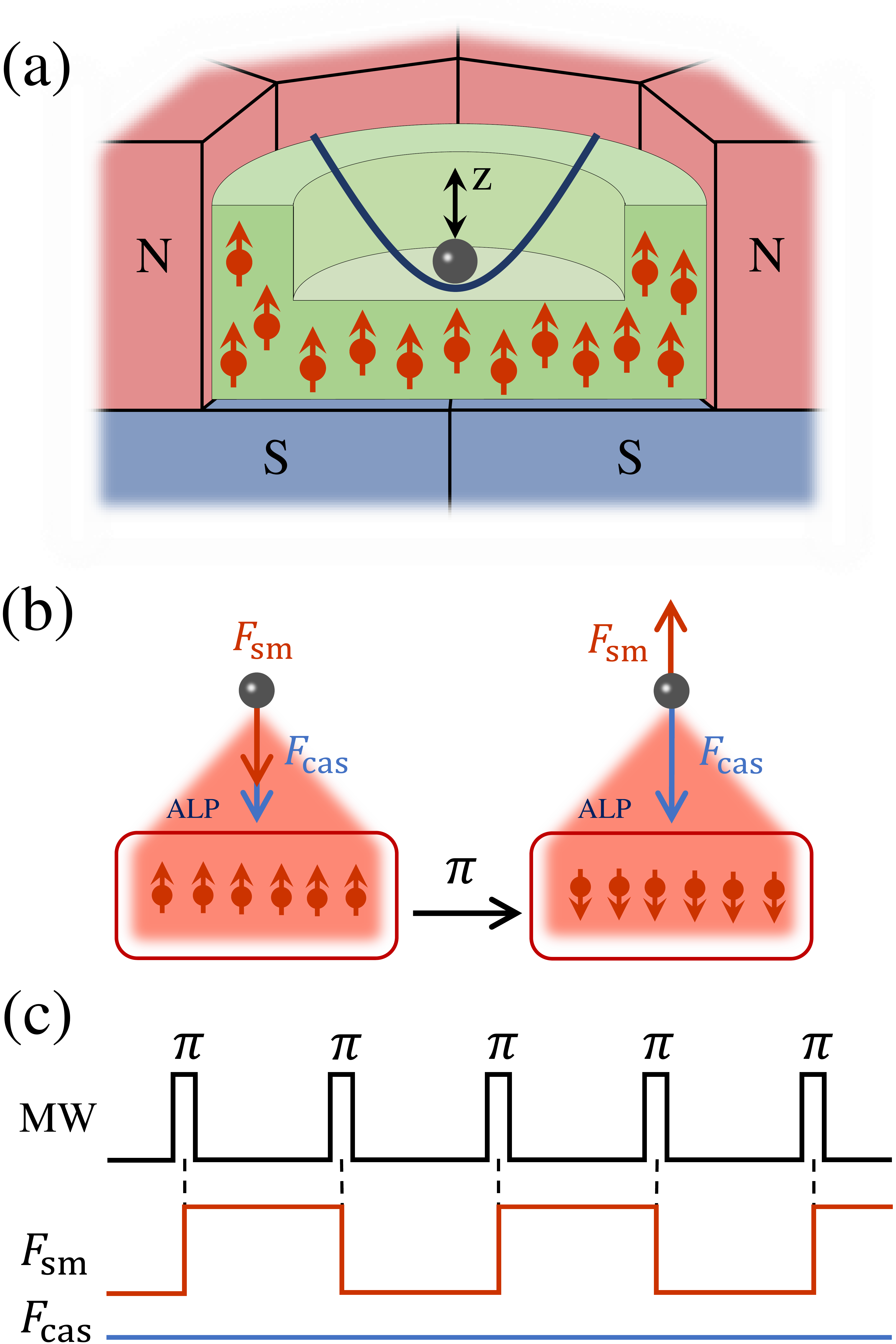}
\caption{(a) Schematic of the proposed experimental system. The red and blue parts of the $\mathrm N$ and $\mathrm S$ poles represent the profile of the permanent magnet, and the green part represents the profile of the spin source. We use a microsphere as the force sensor, which is placed in the magneto-gravitational trap above the surface of the spin source (see Appendix~\ref{sec:A} for the description of its motion). The geometry is sophisticated and is designed to eliminate the spin-induced magnetic force on the levitated microsphere (see Appendix~\ref{sec:B} for details).
 (b) Flipping of the electron-spins by a microwave $\pi$. The spin mass force $F_{\rm sm}$ flips with these spins, while those spin-independent forces, for example the Casimir force $F_{\rm cas}$, are independent of the spins and therefore do not flip with the spins. (c) Microwave (MW) pulse sequences. The spins flip at $2\omega_z$, twice the resonance frequency of the levitated oscillator. This leads to a periodical force $F_{\rm{sm}}$ of the frequency $\omega_{\rm z}$, while the spin-independent forces such as Casimir force $F_{\rm{cas}}$ remains constant during the measurement.}
\label{fig1}
\end{figure}


We use a levitated diamagnetic microsphere mechanical oscillator to investigate the spin-mass interaction [Fig.~\ref{fig1}\textbf{(a)}]. The microsphere is trapped in the magneto-gravitational trap and levitated stably in high vacuum. The diamagnetic-levitated micromechanical oscillator achieves the lowest dissipation in micromechanical and nanomechanical systems to date \cite{Leng_PRApp_2021}, which indicates potentially better force sensitivity than current reported methods. The cryogenic diamagnetic-levitated oscillator described here is applicable to a wide range of mass, making it a good candidate for measuring force with ultra-high sensitivity \cite{Leng_PRApp_2021}.  The position of the microsphere is mainly determined by the equilibrium between the gravity force and the main magnetic force of the trap. A uniform magnetic field is applied to tune the levitation position (see Appendix~\ref{sec:A}).  A groove-shaped electron spin ensemble  (see Appendix~\ref{sec:B} for detail) is located below the mass source as a spin source.

The spin-mass interaction between a polarized electron and an unpolarized nucleon is \cite{Moody_PhysRevD_1984}:\\
\begin{align}
\label{spin mass}V\left(\hat{\sigma}, r\right)=\frac{\hbar^{\rm{2}} g_{\rm{s}}^{\rm{N}} g_{\rm{p}}^{\rm{e}}}{8 \pi m_{\rm{e}}}\left(\bm{\hat{\sigma}\cdot e_{\rm{r}}}\right)\left(\frac{1}{\lambda r}+\frac{1}{r^{\rm{2}}}\right)e^{\rm{-r/\lambda}},
\end{align}
where $g_{\rm{s}}^{\rm{N}}$ and $g_{\rm{p}}^{\rm{e}}$ are the coupling constants of the interaction, with $g_{\rm{s}}^{\rm{N}}$ representing the axion scalar coupling constant to an unpolarized nucleon and $g_{\rm{p}}^{\rm{e}}$ representing the axion pseudoscalar coupling constant to an electron spin, $\lambda= \hbar/(m_{\rm{a}} c)$ is the interaction range, $m_{\rm{a}}$ is the ALP mass, $m_{\rm{e}}$ is the mass of electron, $\bm{\hat{\sigma}}$ is electron spin operator, $r$  is the displacement between the electron and nucleon, and $\bm{e_{\rm{r}}}$ \ is the direction. The spin-mass force along $z$ axis is calculated by integrating the force element between microsphere and spin source based on Eq.\eqref{spin mass} as:
\begin{align}
\label{fsm}
F_{\rm sm}(t)=\rho_{\rm{e}}\left(t\right)\frac{\hbar^2g_{\rm{s}}^{\rm{N}}g_{\rm{p}}^{\rm{e}}\rho_{\rm{m}}}{8\pi m_{\rm{e}}}\zeta_{\rm sm}(R,d,\lambda),
\end{align}
where $\rho_{\rm{e}}\left(t\right)$ is the time-dependent net electron spin density along $z$ axis, $\rho_{\rm m}$ is the nucleon density of the microsphere, $\zeta_{\rm sm}(R,d,\lambda)$ is the effective volume for spin-mass interaction that depends on geometry parameters (see Appendix~\ref{sec:C}), $R$ is the microsphere radius, and $d$ is the surface distance between the mass and the spin source.

The electrons' spins are initially polarized along the magnetic field under high field and low temperature so that $\rho_{\rm{e}}\left(0\right)=\rho_{\rm{e0}}$, where $\rho_{\rm{e0}}$ is the electron density
of the spin source. Then they are flipped periodically in resonance with the microsphere mechanical oscillator [see Fig.~\ref{fig1}\textbf{(b)}]. On one hand, the spin-independent interactions, such as the Casimir force,  will be off-resonance and become eliminated [Fig.~\ref{fig1}\textbf{(c)}]. On the other hand, the spin-mass interaction is preserved on the resonance condition. The spin autocorrelation function is defined as $\left\langle\rho_{\rm{e}}\left(t\right)\middle|\rho_{\rm{e}}\left(0\right)\right\rangle=\rho_{\rm{e}}(0)^{2}P\left(t\right)=\rho_{\rm{e}}(0)^{2}e^{\rm{-t/T_1}} \xi(t)$, where $T_{\rm{1}}$ is the electron spin-lattice relaxation time and $\xi(t)$ is the modulation function (see Appendix~\ref{sec:D}). The microwave $\pi$ pulses flip the electron spins periodically with frequency $2\omega_z$. $\xi(t)$ jump between $-1$ and $+1$ every time the electron spins are flipped. The corresponding power spectral density (PSD) of the spin-related force is proportional to $\widetilde{G}\left(\omega\right)$, which is the Fourier transform of $P(t)$.  The PSD of spin-mass force is then :
\begin{align}
S_{\rm{ff}}^{\rm{sm}}\left(\omega\right)=\left(\frac{\hbar^2g_{\rm{s}}^{\rm{N}}g_{\rm{p}}^{\rm{e}}\rho_{\rm{m}}}{8\pi m_{\rm{m}}}\zeta_{\rm sm}(R,d,\lambda)\right)^2\rho^2_{\rm{e_0}}\widetilde{G}\left(\omega\right) .
\label{Ssm}
\end{align}
If the spin-mass interaction signal is observed on resonance ($\omega=\omega_{\rm z}$), the coupling $g_{\rm s}^{\rm N}g_{\rm p}^{\rm e}$ can be derived as
\begin{align}
g_{\rm s}^{\rm N}g_{\rm p}^{\rm e}=\sqrt{\frac{S_{\rm ff}^{\rm sm}(\omega_{\rm z})}{\widetilde{G}(\omega_{\rm z})}}\frac{8\pi m_{\rm e}}{\zeta_{\rm sm}\hbar^2\rho_{\rm m}\rho_{\rm e_0}}.
\end{align}

Apart from the spin-mass force, spin-induced magnetic force $F_{\rm{s}}$ between electron spins and the diamagnetic microsphere is recorded during the measurement. Fortunately, well designed spin-source geometry can eliminate most of the force (see Appendix~\ref{sec:B}). Then the residual spin-induced magnetic force is
\begin{align}
F_{\rm{s}}(t) = \rho_{\rm{e}}(t)\frac{\mu_{\rm{B}}\chi_{\rm{m}}}{2}\frac{\partial B_{\rm{0z}}}{\partial z}\zeta_{\rm{s}}(R,d),
\end{align}
where $\zeta_{\rm s}(R,d)$ is the effective volume for spin-induced force. Similarly, the PSD of $F_s$ is
\begin{align}
S_{\rm{ff}}^{\rm{s}}\left(\omega\right)=\left(\frac{1}{2}\mu_{\rm{B}}\chi_{\rm{m}}\frac{\partial B_{\rm{0z}}}{\partial z}\zeta_{\rm{s}}(R,d)\right)^2\rho^2_{\rm{e_0}}\widetilde{G}\left(\omega\right).
\label{Ss}
\end{align}

Considering the fluctuating noise, the equation of motion for the system center of mass is
\begin{align}
m\ddot{z}+m\gamma\dot{z}+ m \omega_{\rm{z}}^2 z=
F_{\rm{flu}}\left(t\right)+F_{\rm{s}}(t)+F_{\rm{sm}}(t),
\end{align}where $m$ is the mass of the microsphere, $\omega_{\rm z}/2\pi$ is the resonance frequency, $\gamma/2\pi$ is the intrinsic damping rate, and $F_{\rm flu}(t)$ is the fluctuating noise force that includes the thermal Langevin force $F_{\rm th}(t)$ and the radiation pressure fluctuations $F_{\rm ba}(t)$ \cite{Clerk_2010_RMP}.

The total detected displacement PSD is given by:
\begin{align}
S_{\rm{zz}}^{\rm{tot}}(\omega)=S_{\rm{zz}}^{\rm{imp}}(\omega)+\frac{{|\chi\left(\omega\right)|}^2}{m^2}\left(S_{\rm{ff}}^{\rm {ba}}+S_{\rm{ff}}^{\rm{th}}+S_{\rm{ff}}^{\rm{s}}+S_{\rm{ff}}^{\rm{sm}}\right)
\label{S_noise}
\end{align}
where $\chi\left(\omega\right)$ is the mechanical susceptibility given by $\left|\chi\left(\omega\right)\right|^2=1/[\left(\omega_{\rm{z}}^2-\omega^2\right)^2+\gamma^2\omega^2$];  $S_{\rm{zz}}^{\rm{imp}}(\omega)$ denotes the PSD of the detector imprecision noise; $S_{\rm{ff}}^{\rm {ba}}$, $S_{\rm{ff}}^{\rm {th}}$, $S^{\rm{s}}_{\rm {ff}}$, and $S_{\rm{ff}}^{\rm {sm}}$ are the PSDs of $F_{\rm ba}(t)$, $F_{\rm th}(t)$, $F_{\rm s}(t)$, and $F_{\rm sm}(t)$, respectively. The total fluctuation noise $S_{\rm ff}^{\rm flu}(\omega_{\rm z})=S_{\rm ff}^{\rm th}+S_{\rm ff}^{\rm ba}+m^2S_{\rm zz}^{\rm imp}{|\chi(\omega_{\rm z})|}^{-2}$. Due to these noises, the detection limit of spin-mass coupling strength $g_{\rm s}^{\rm N}g_{\rm p}^{\rm e}$ is thus:
\begin{align}
\left(g_{\rm s}^{\rm N}g_{\rm p}^{\rm e}\right)_{\rm limit}=\sqrt{\frac{S_{\rm ff}^{\rm flu}(\omega_{\rm z})+S_{\rm ff}^{\rm s}(\omega_{\rm z})}{\widetilde{G}(\omega_{\rm z})}}\frac{8\pi m_{\rm e}}{\zeta_{\rm sm}\hbar^2\rho_{\rm m}\rho_{\rm e_0}}.
\label{gsgp}
\end{align}

\section{Estimated Results}

Here we propose a reasonable scheme by our estimation. We consider a microsphere with mass $m=1.5\times10^{-13}$kg and radius $R=3.2~\rm{\mu m}$ of density $1.1\times10^{3}~\rm{kg/m^{-3}}$. Thus, the corresponding nucleon density is $\rho_{m}=6.7\times10^{29}m^{-3}$. The magnetic susceptibility of the microsphere is $-$$9.1\times10^{-6}$. The whole system is proposed to be placed in a cryostat with temperature $T$= 20 mK and external uniform magnetic field $B_{\rm ext}=1.85$ T. A permanent magnet provides 0.15 T magnetic field and correspondingly the $z$ direction magnetic gradient ${\partial B_{0z}}/{\partial z}=750$ T/m. The microsphere is then levitated with a surface distance $d=1.46~\mu$m above the spin source. The whole mechanical oscillator system has a typical frequency of the $z$ axis \cite{Leng_PRApp_2021,ZhengDi_PhysRevResearch_2020}, and the electron density of the spin source is $\rho_{e_0}=2.3\times10^{27}$ m$^{-3}$. The direction of the electron spins is initially prepared along the external magnetic field, which in our design is approximately along the $z$ axis, with a maximum tilted angle  of 4$^{\circ}$.

\begin{figure}[hbtp]
\includegraphics[width=8cm]{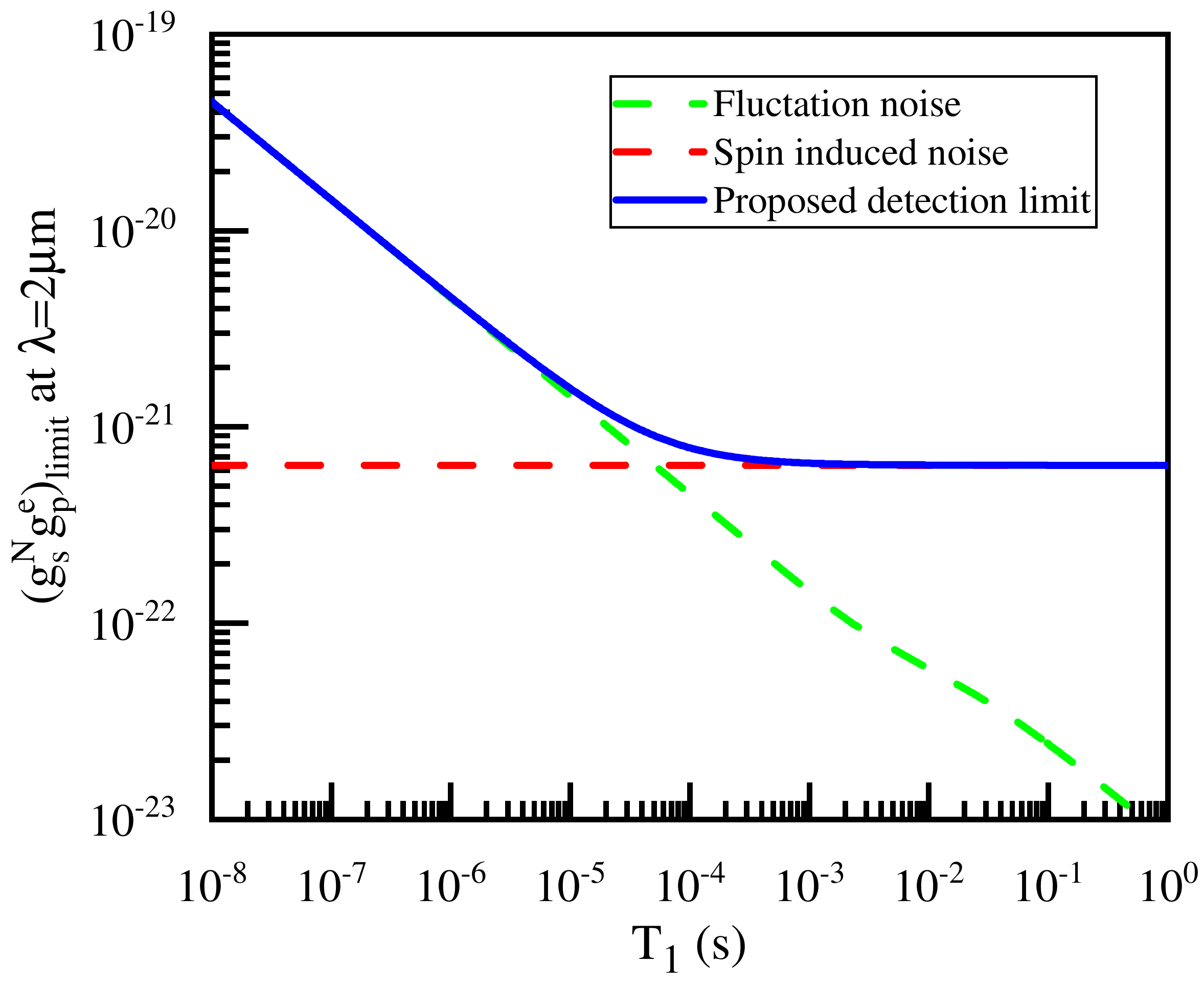}
\caption{$(g_{\rm s}^{\rm N}g_{\rm p}^{\rm e})_{\rm limit}$  for the force range of ${\rm \lambda=2\mu m}$ as an example. The green line denotes $(g_{\rm s}^{\rm N}g_{\rm p}^{\rm e})_{\rm limit}$ calculated from the total fluctuation noise, which decreases as $T_1$ increases. The red line denotes the correction of $(g_{\rm s}^{\rm N}g_{\rm p}^{\rm e})_{\rm limit}$ by taking the residual spin-induced magnetic force into account, which is independent of $T_1$. The blue curve denotes their sum. The correction from spin-induced magnetic force (red curve) is dominant when $T_1>1~$ms.}
\label{GsGp_T1}
\end{figure}

The total measurement time is set as 1s. We take the proposed experimental sensitivity limited by the total fluctuation noise as  $
S_{\rm ff}^{\rm flu}(\omega_{\rm z})=S_{\rm ff}^{\rm th}+S_{\rm ff}^{\rm ba}+m^2S_{\rm zz}^{\rm imp}{|\chi(\omega_{\rm z})|}^{-2}$. Here
$S_{\rm ff}^{\rm th} $ is estimated to be $5.14\times10^{-43}{\rm N^2/Hz}$ according to $S_{\rm ff}^{\rm th} = 4m \gamma k_{\rm B} T$, with $\gamma/2\pi = 10^{-6}$Hz \cite{Leng_PRApp_2021}.
Imprecision noise and backaction noise are related, when they contribute equally the sum has a minimum $S_{\rm ff}^{\rm sum}(\omega_{\rm z})=S_{\rm ff}^{\rm ba}+m^2S_{\rm zz}^{\rm imp}{|\chi(\omega_{\rm z})|}^{-2}=2m{|\chi(\omega_{\rm z})|}^{-1}\hbar/\eta^{\frac{1}{2}}$. In a practical condition, the measurement efficiency $\eta \geq 0.001$  \cite{Tebbenjohanns_2019_PRL}, which implies $S_{\rm ff}^{\rm sum}(\omega_{\rm z})=9.36\times10^{-49}{\rm N^2/Hz}$. Thus, the total fluctuation noise is dominated by the thermal noise, with $S_{\rm ff}^{\rm flu}(\omega_{\rm z})\approx5.14\times10^{-43}{\rm N^2/Hz}$. Under such an estimated experimental sensitivity, $(g_{\rm s}^{\rm N}g_{\rm p}^{\rm e})_{\rm limit} = 8\pi m_{\rm e}[S_{\rm ff}^{\rm flu}(\omega_{\rm z})/\widetilde{G}(\omega_{\rm z})]^{\frac{1}{2}}/\zeta_{\rm sm}\hbar^2\rho_{\rm m}\rho_{\rm e_0}$. As $\widetilde{G}(\omega_{\rm z})$ is proportional to the electron spin-lattice relaxation time, $(g_{\rm s}^{\rm N}g_{\rm p}^{\rm e})_{\rm limit}$ decreases as $T_1$ increases, which is  shown in green in Fig.~\ref{GsGp_T1}.\\

Practically, it is not feasible to completely eliminate the spin-induced magnetic force due to fabrication imperfection of the spin-source geometry (see Appendix~\ref{sec:B}). A correction for $(g_{\rm s}^{\rm N}g_{\rm p}^{\rm e})_{\rm limit}$ is introduced as follows. Since the spin-induced magnetic noise is spin-dependent while the $\widetilde{G}(\omega_{\rm z})$ has the same scaling, its contribution to $(g_{\rm s}^{\rm N}g_{\rm p}^{\rm e})_{\rm limit}$ is constant (blue curve in Fig.~\ref{GsGp_T1}). For $T_1>1{\rm ms}$, the $(g_{\rm s}^{\rm N}g_{\rm p}^{\rm e})_{\rm limit}$ is dominated by the spin-induced magnetic force and approaches to the minimum  $8\pi m_{\rm e}[S_{\rm ff}^{\rm s}(\omega_{\rm z})/\widetilde{G}(\omega_{\rm z})]^{\frac{1}{2}}/\zeta_{\rm sm}\hbar^2\rho_{\rm m}\rho_{\rm e_0}$ under our estimation.

Finally, Fig.~\ref{fig3} shows the calculated $(g_{\rm{s}}^{\rm{N}}g_{\rm{p}}^{\rm{e}})_{\rm limit}$ (see Appendix~\ref{sec:E}) set by the proposed experiment at $\lambda=0.5\rm{\mu m}$ -- 50 $\rm{\mu m}$ together with reported experimental results for the constraints of spin-mass coupling. Here our result is estimated through supposing $T_1=1$s, for spin-lattice relaxation the time can be longer than the scale of seconds at low temperature~\cite{Reynhardt_1998_JCP, Takahashi_2008_PRL}. The limitation for our proposal is the residual spin-induced magnetic force, which can not be eliminated by spin flip procedures.
For $\lambda=2~\rm{\mu m}$, the minimum detectable spin-mass coupling is $(g_{\rm{s}}^{\rm{N}}g_{\rm{p}}^{\rm{e}})_{\rm limit}=4.3\times 10^{\rm{-22}}$ (~\autoref{tableSff}), by our estimation due to the spin-induced magnetic noise $S_{\rm ff}^{\rm s}$ under reasonable fabrication imperfection $\Delta \zeta_{\rm{s}}= 1.38\times 10^{\rm{-21}}~\rm{m^3}$ (see Appendix~\ref{sec:B}). In conclusion, compared to those from Ref.  \cite{Stadnik_PRL_2018,Hoedl_PhysRevLett_2011}, our result shows an improvement of nearly 3 orders of magnitude more stringent at the ALP mass range of $7~\rm{meV}$.

\begin{figure}
\includegraphics[width=8.6cm]{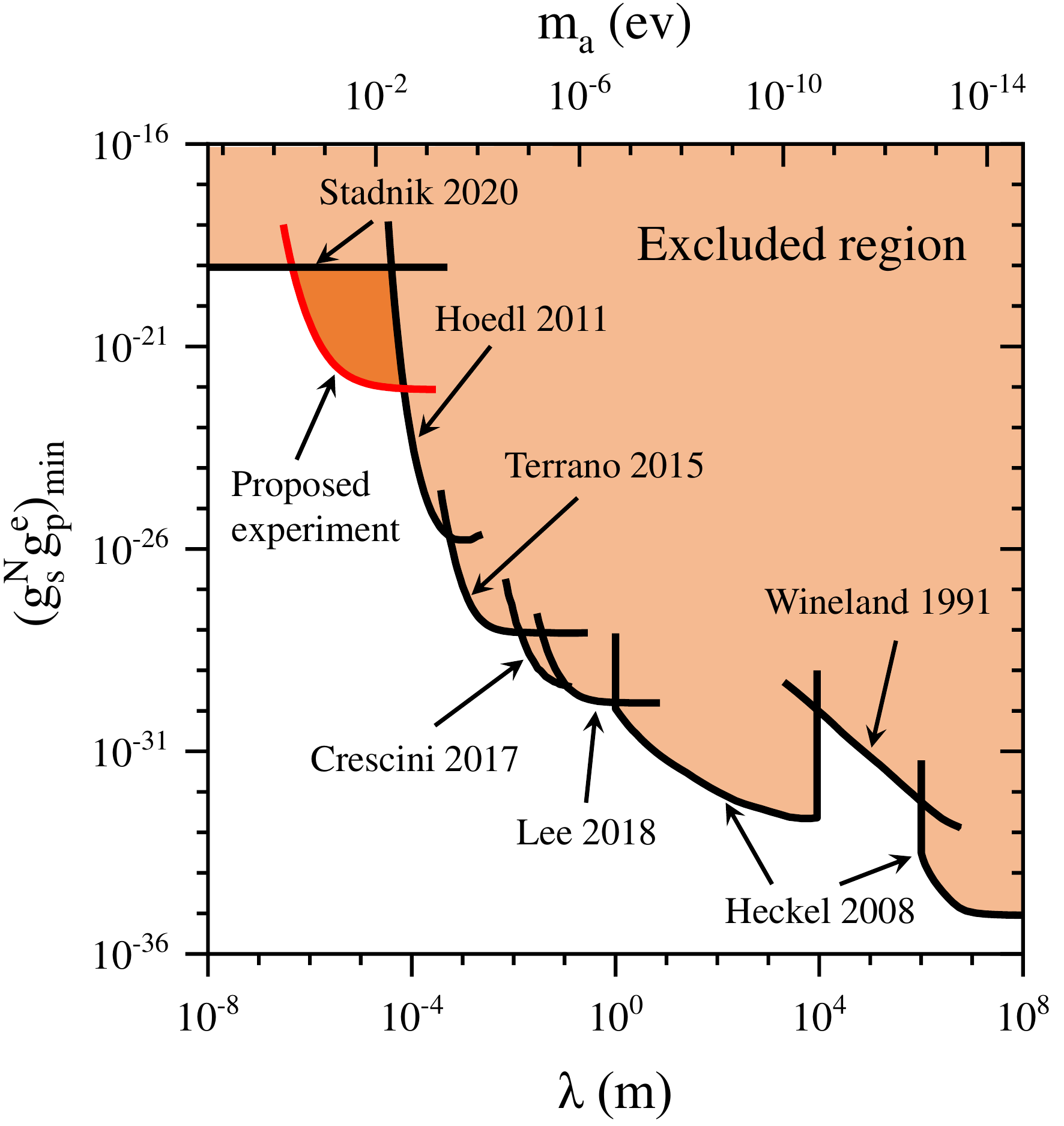}
\caption{$(g_{\rm{s}}^{\rm{N}}g_{\rm{p}}^{\rm{e}})_{\rm limit}$ as a function of the interaction range $\lambda$. Black solid lines represent the results from Refs.\ \cite{Wineland_PhysRevLett_1991,Heckel_PhysRevD_2008,Hoedl_PhysRevLett_2011,Terrano_PhysRevLett_2015,Crescini_PhysLetbB_2017,Lee_PhysRevLett_2018,Stadnik_PRL_2018}. The estimated bound of our method is plotted in red for the spin-mass force range $\lambda=0.5~\mu$m -- $50~\mu$m. Our result is nearly 3 orders of magnitude more stringent at the ALP mass of $7~\rm{meV}$ compared with those from Ref.  \cite{Stadnik_PRL_2018,Hoedl_PhysRevLett_2011}.}
\label{fig3}
\end{figure}

\section{DISCUSSION}
The nearly 3 orders of magnitude enhancement in our scheme comes from the following two aspects. First, the magnetic resonance spin flipping is applied to suppress the short-range force noise which limits the precision of probing spin-mass coupling.
Second, the diamagnetic levitation realizes an ultra-low dissipation in comparison with other reported mechanical systems, and this together with a low-temperature condition provides an ultralow detection noise. The main limitation of our method comes from the spin-induced magnetic force that evolves in accord with the spin-mass interaction, which cannot be eliminated  with finite size of the force sensor and imperfect geometric symmetry in the layout of the electron spins. Such a magnetic background could be measured by a sensitive magnetometer with high spatial resolution, such as a single nitrogen-vacancy (NV) center, and then be subtracted from the measured signal, leading to more stringent constraints of the spin-mass coupling strength $g_{\rm{s}}^{\rm{N}}g_{\rm{p}}^{\rm{e}}$. Other technical sources of noise or systematics may come from the heating effect and experimental system vibration. Under high vacuum and low temperature, the microsphere is likely to be heated by the continuous illumination light, leading to a higher effective center-of-mass temperature than the sample chamber. Such a heating effect can be suppressed by pulsed detection light. The system vibration can be eliminated further when the system noise is already very small (under 20 mK).

\begin{table*}[htbp]
	\centering
\caption{Contributions to the power spectral density (PSD)}
	\begin{tabular}{ccc}
        \botrule
        PSD calculated at $T_1=1$s &\hspace{1cm} Size ($N^2/$Hz)  & \hspace{1cm}$\rm Contribution\ to\ (g_{\rm{s}}^{\rm{N}}g_{\rm{p}}^{\rm{e}})_{\rm min}$ at $\lambda=2~\mu$m    \\ \colrule
		$\rm Of\ spin\  induced\ magnetic\ force\ S_{\rm ff}^{\rm s}(\omega_{\rm z})$   &\hspace{1cm}$2.59\times10^{-41}$  &\hspace{1cm}$4.3\times10^{-22}$      \\
        $\rm Of\ thermal\ noise\ S_{\rm ff}^{\rm th}(\omega_{\rm z})$   &\hspace{1cm}$5.14\times10^{-43}$  &\hspace{1cm}$9.1\times10^{-24}$      \\
        $\rm Of\ backaction\ noise\ plus\ imprecision\ noise\ S_{\rm ff}^{\rm add}(\omega_{\rm z})$   &\hspace{1cm}$9.36\times10^{-49}$  &\hspace{1cm}$1.2\times10^{-26}$      \\
        $\rm Total$   &\hspace{1cm}$2.59\times10^{-41}$  &\hspace{1cm}$4.3\times10^{-22}$      \\
		\botrule
	\end{tabular}
\label{tableSff}
\end{table*}

\begin{acknowledgments}
This work was supported by the National Key R\&D Program of China (Grant No.~2018YFA0306600), the National Natural Science Foundation of China (Grant No.~61635012, No.~11675163, No.~11890702, No.~12075115, No.~81788101, No.~11761131011, No.~11722544 and No.~ U1838104),
the CAS (Grant No.~QYZDY-SSW-SLH004, No.~GJJSTD20170001, and No.~Y95461A0U2), the Fundamental Research Funds for the Central Universities (Grant No.~021314380149), and the Anhui Initiative in Quantum Information Technologies (Grant No.~AHY050000).
\end{acknowledgments}

\appendix
\section{DYNAMICS OF MICRO-SPHERE OSCILLATOR}\label{sec:A}
\begin{figure}[htp]
\includegraphics[width=8cm]{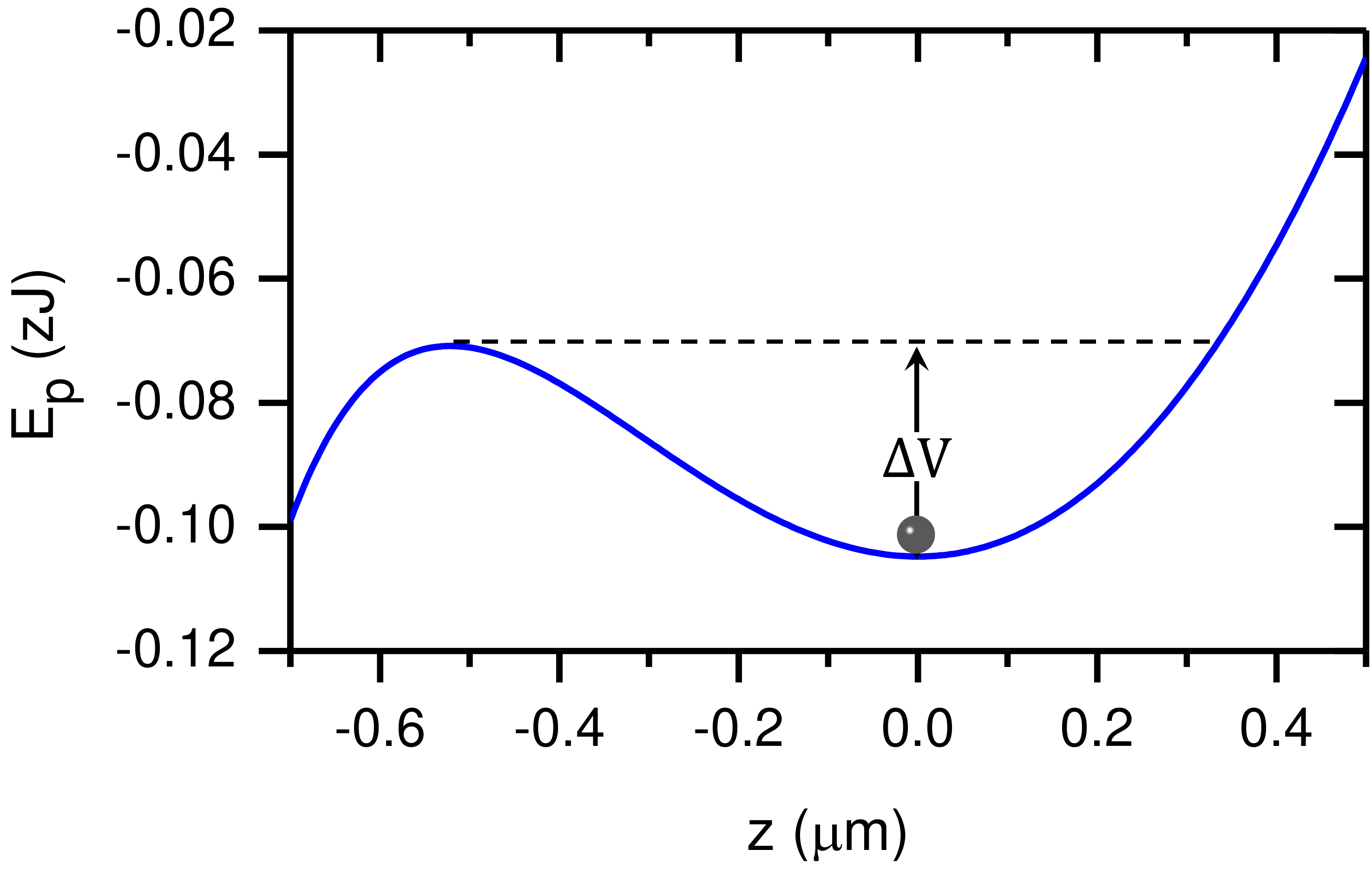}
\caption{The magneto-gravity potential $E_{\rm p}$ as a function of the displacement of the microsphere. The depth of the trap is $\Delta V = 3.4\times10^{-23}J$. According to the energy equipartition theorem, the thermal energy at $20~$mK is $1.38\times10^{-25}J$ , which implies that the microsphere will not escape the trap. }
\label{Potential}
\end{figure}

For the microsphere, the dynamics in the $z$ direction of its center of mass (CM) in our system reads:
\begin{align}
m\ddot{z}=-m\gamma \dot{z}+F_{\rm sm}+F_{\rm s}+F_{\rm flu}+\frac{-\partial E_{\rm p}}{\partial z},
\label{mz}
\end{align}
where $m\gamma \dot{z}$ is the residual air damping force, $F_{\rm sm}$ is spin-mass force, $F_{\rm s}$ is the spin-induced magnetic force, $F_{\rm flu}$ is the fluctuating noise force. $E_{\rm p}$ is the trap potential subject to gravitational field, main magnetic field, spin-induced magnetic field, and Casimir attractive force, i.e.,
\begin{align}
E_{\rm p}=mgz+\int_{\rm m}{\rm d V}\frac{\chi_{\rm m}}{2\mu_0}B^2_{\rm 0z}+\int_{\rm m}{\rm d V}\frac{\chi_{\rm m}}{2\mu_0}B^2_{\rm sz}+V_{\rm cas},
\label{Ep}
\end{align}
where $mgz$ is the gravity of microsphere, $\int_{\rm m}{\rm d V}$ represents the volume integral over the microsphere, $\mu_0$ is permeability of vacuum, and $\chi_{\rm m}$ is magnetic susceptibility of the microsphere. $B_{\rm 0z}$ is the main magnetic field at the CM of the microsphere, which is the sum of the magnetic field generated by permanent magnet and the uniform external magnetic field $B_{\rm ext}$. $B_{\rm sz}$ accounts for the spin-induced magnetic field, and $V_{\rm cas}$ is the Casimir potential \cite{Klimchitskaya_PRA_2000,Bordag_PR_2001,Decca_PRL_2003} between the surface of microsphere and the surface of spin source, reads as,
\begin{align}
V_{\rm cas}=-\frac{\hbar c\pi^2}{1440(z+d)^2}2\pi R\eta_{\rm c},
\label{cas}
\end{align}
where $R$ is the radius of the microsphere, $z$ corresponds to the displacement of the microsphere, $d = 1.46 {\rm ~\mu m}$ is the surface distance between the microsphere and the spin source when the microsphere locates in equilibrium  (Fig.~\ref{Potential}), and $\eta_{\rm c}=0.059$ characterizes the reduction in the Casimir force, depending on the dielectric functions of the microsphere and the spin source. The value of $E_{\rm p}$ versus the displacement of the microsphere is shown in Fig.~\ref{Potential}.

Thus our mechanical system can be described
 as a damping harmonic oscillator subject to $F_{\rm sm}$, $F_{\rm s}$ and $F_{\rm flu}$, i.e.,
\begin{align}
m\ddot{z}+m\gamma \dot{z}+m\omega^2_{\rm z}z=F_{\rm sm}+F_{\rm s}+F_{\rm flu},
\end{align}
where $\omega_{\rm z}$ is the resonant frequency of the microsphere,
\begin{align}
\omega_{\rm z}=\sqrt{\frac{1}{m}\frac{\partial^2 E_{\rm p}}{\partial z^2}}.
\end{align}

The equilibrium position of the microsphere can be derived by $\partial E_{\rm p}/\partial z=0$. The spin-induced magnetic field and $V_{\rm cas}$ are so weak that they have negligible influence on this trap, so that the equilibrium position is mainly determined by the gravity field  and the main magnetic field $B_{\rm 0z}$. Thus, we can indirectly tune it by  the uniform external magnetic field $B_{\rm ext}$.

\section{DESCRIPTION OF PROPOSED EXPERIMENT SCHEME}\label{sec:F}

The proposed experiment setup is based on a microsphere made of polyethylene glycol levitated in a magneto-gravitational trap similar to that demonstrated in Ref. \cite{Leng_PRApp_2021}. The microsphere is first
neutralized using an ultraviolet light, the magneto-gravitational trap is placed in the high vacuum cryostat and a spring-mass suspension system is used to isolate external vibration noise. A weak laser is focused on the microsphere and scattering light is collected via lens. A CMOS camera is used to record position of the microsphere, and an avalanche photo-diode detector (APD) is used to measure the dynamics of the oscillator.
To control and realize feedback cooling of the oscillator, due to the low frequency and long coherent time, a program based feedback is used to generate feedback current signal $I$, which will generate
small magnetic field, combined with the magnetic field gradient of the magneto-gravitational trap, and the feedback force needed to cool the mode motional temperature is realized. The spin flip control is realized by electron spin resonance  technique. A  microwave pulse with a resonant frequency $f=\gamma_{\rm e}B_{\text{ext}}$ (where $\gamma_{\rm e}$ is gyromagnetic ratio of electron) is applied to the electron spin source through a coplanar waveguide to drive the spin state. The $\pi$ pulse with time length $t=1/2\gamma_eB_1$ flips the electron spin from $|\uparrow\rangle$  to $|\downarrow\rangle$ or vice versa.

\section{AUTOCORRELATION FUNCTION OF NET ELECTRON-SPINS DENSITY}\label{sec:D}
\begin{figure}[htp]
\includegraphics[width=8.6cm]{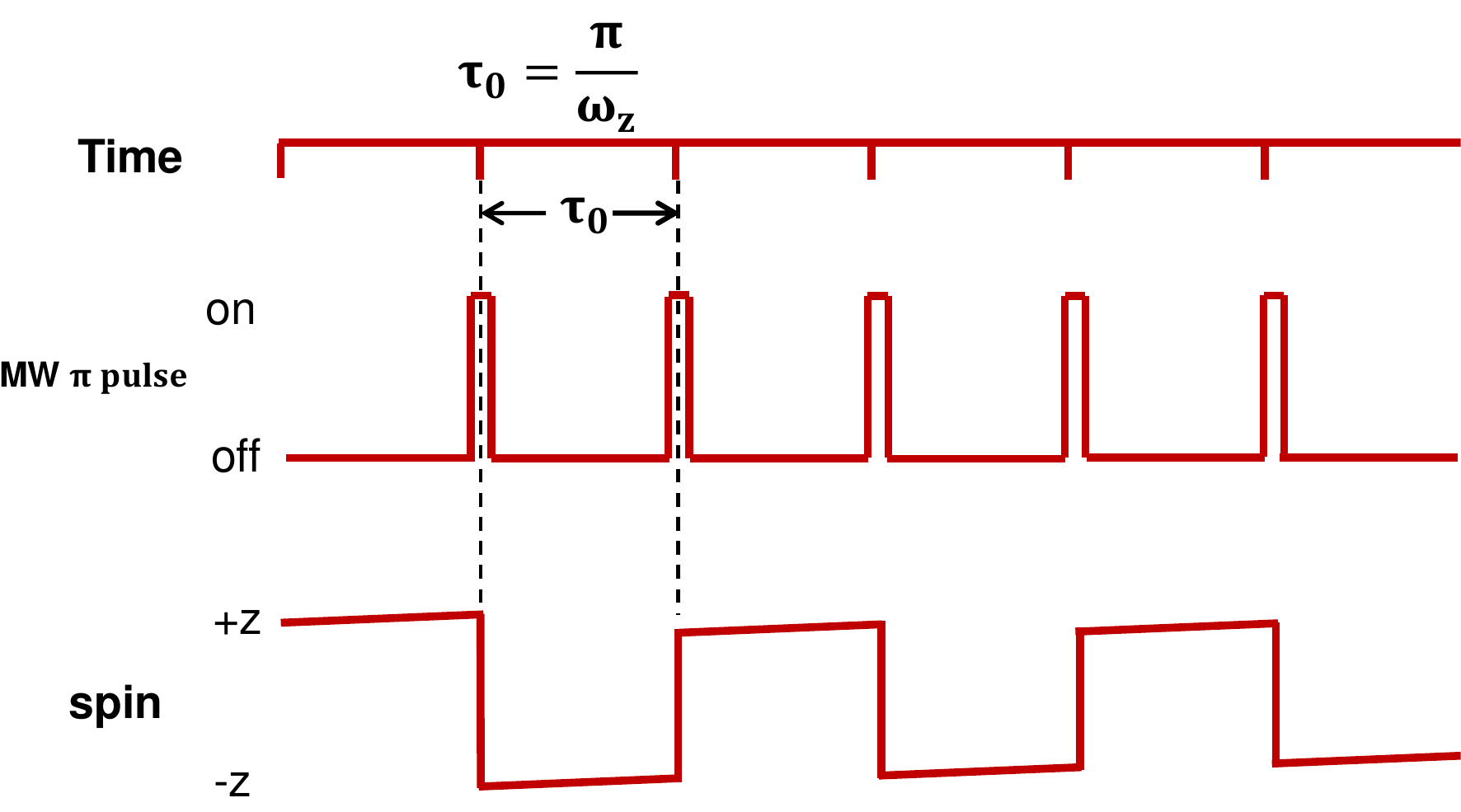}
\caption{Microwave $\pi$ pulses carried on with a frequency of $2\omega_{\rm{z}}$. The time interval between two adjacent $\pi$ pulses is $\tau_{\rm 0}=\pi/\omega_{\rm{z}}$. Spin flips with a frequency of $\omega_{\rm{z}}$, and its amplitude varies slowly over time due to the spin-lattice relaxation.}
\label{PiPulse}
\end{figure}

The autocorrelation function of electron polarization is $\langle\rho_{\rm e}(t)\rho_{\rm e}(0)\rangle$. Suppose these electron spins are independent of each other, we have
\begin{align}
 \langle\rho_{\rm e}(t)\rho_{\rm e}(0)\rangle=\rho^2_{\rm e0}P(t),
\end{align}
where $P(t)$ is the autocorrelation function of a single spin, i.e.,
\begin{align}
P(t)=p_{\rm\uparrow}(t)-p_{\rm\downarrow}(t).
\end{align}
Here ${ p_{\uparrow,\downarrow}}(t)$ represents the spin population on ${\rm|\uparrow\rangle}$ or ${\rm|\downarrow\rangle}$. Every time
a $\pi$ pulse is applied to flip the electron spin,
\begin{align}
\notag
p_{\rm\uparrow}(t)=p_{\rm\uparrow}(t)+p_{\rm\downarrow}(t)p_1(\tau)\\
p_{\rm\uparrow}(t)=p_{\rm\uparrow}(t)[1-p_1(\tau)],
\label{P_}
\end{align}
where $\tau\in(0, \tau_{\rm 0})$, $\tau_0$ corresponds to the period between two adjacent $\pi$ pulses (Fig.~\ref{PiPulse}), $p_1(\tau)=1-e^{\rm -\tau/T_1}$ is the spin-flip probability during $\tau_0$, ${ T_1}$ is spin-lattice relaxation time.

\begin{figure}[htp]
\includegraphics[width=8cm]{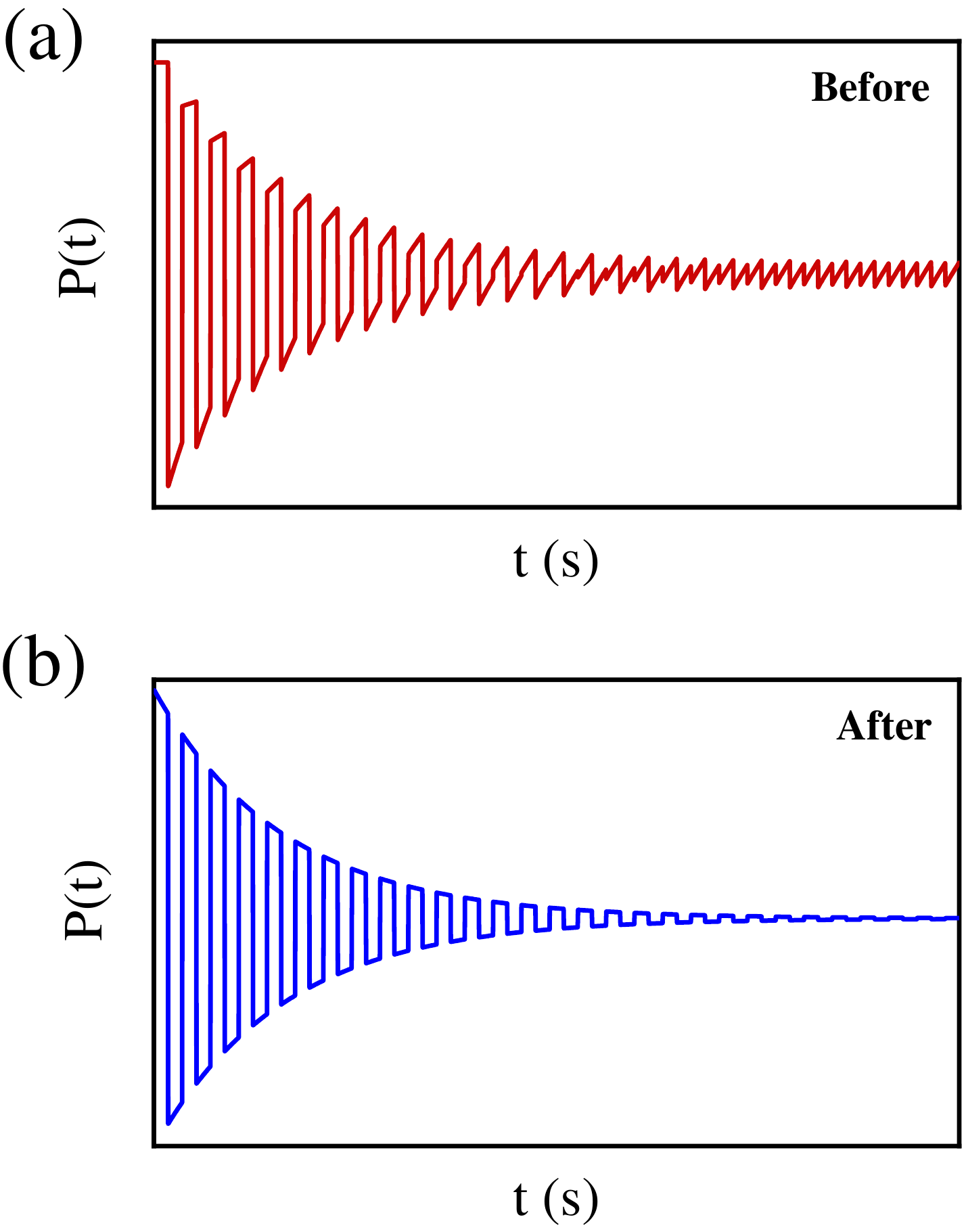}
\caption{(a)\ Variation of the native P(t) after equilibrium in the form of a sawtooth-like wave. The frequency of the sawtooth-like wave is $2\omega_{\rm z}$, which is out of resonance with the microsphere oscillator and can be neglected. (b)\ Effective P(t) after dropping the sawtooth-like wave. It is a square wave modulated signal that decays exponentially with time, which is determined by the spin-lattice relaxation time. The frequency of this signal is $\omega_{\rm z}$.}
\label{P_t}
\end{figure}

The evolution of $P(t)$ is shown in Fig.~\ref{P_t}(a). $P(t)$ presents a sawtooth-like wave of frequency $2\omega_{\rm z}$ for $t = k\tau_0 +\tau \gg T_1$, (k=0, 1, 2, \ldots),
\begin{align}
P(\tau + k \tau_0)=1-\frac{2e^{-\frac{\tau}{T_1}}}{1+e^{-\frac{\tau_0}{T_1}}},~~\tau\in(0, \tau_0)
\end{align}
Only the signal with resonant frequency $\omega_z$ needs to be collected. After dropping the sawtooth-like signal whose frequency is $2\omega_{\rm z}$, the resonant signal is shown in Fig.~\ref{P_t}(b). The resonant signal is a square wave with a exponential decay, i.e.,
\begin{align}
P(t)=e^{\rm{-t/T_1}} \xi(t),
\end{align}
where $\xi(t)$ is the modulation function of the following form
\begin{align}
\xi(t)=\frac{2}{1+e^{-\frac{\tau_0}{T_1}}}\varsigma(\omega_{\rm z}t+\frac{\pi}{2}).
\end{align}
Here $\varsigma(\omega_{\rm z}t+\frac{\pi}{2})$ is a square wave of frequency $\omega_{\rm z}$.
According to the Wiener-Khinchine theorem, its single side PSD is:
\begin{widetext}
\begin{align}
\notag
\widetilde{G}(\omega)&=\frac{4}{1+e^{-\tau_0/T_1}} \left(\frac{2T_1}{1+T_1^2\omega^2}-4 e^{-\tau_0/2 T_1}\frac{T_1\left(1+e^{-\tau_0/ T_1}\right)\cos\left(\omega\tau_0/2\right)-\omega\left(1-e^{-\tau_0/ T_1}\right)\sin\left(\omega\tau_0/2\right)}{\left(1+T_1^2\omega^2\right)\left(1+e^{-2\tau_0/T_1}+2e^{-\tau_0/T_1}  \cos\left(\omega\tau_0\right)\right)}\right).
\end{align}
\end{widetext}

\section{PSD OF SPIN INDUCED MAGNETIC FORCE}\label{sec:B}

\begin{figure}[htp]
\includegraphics[width=6cm]{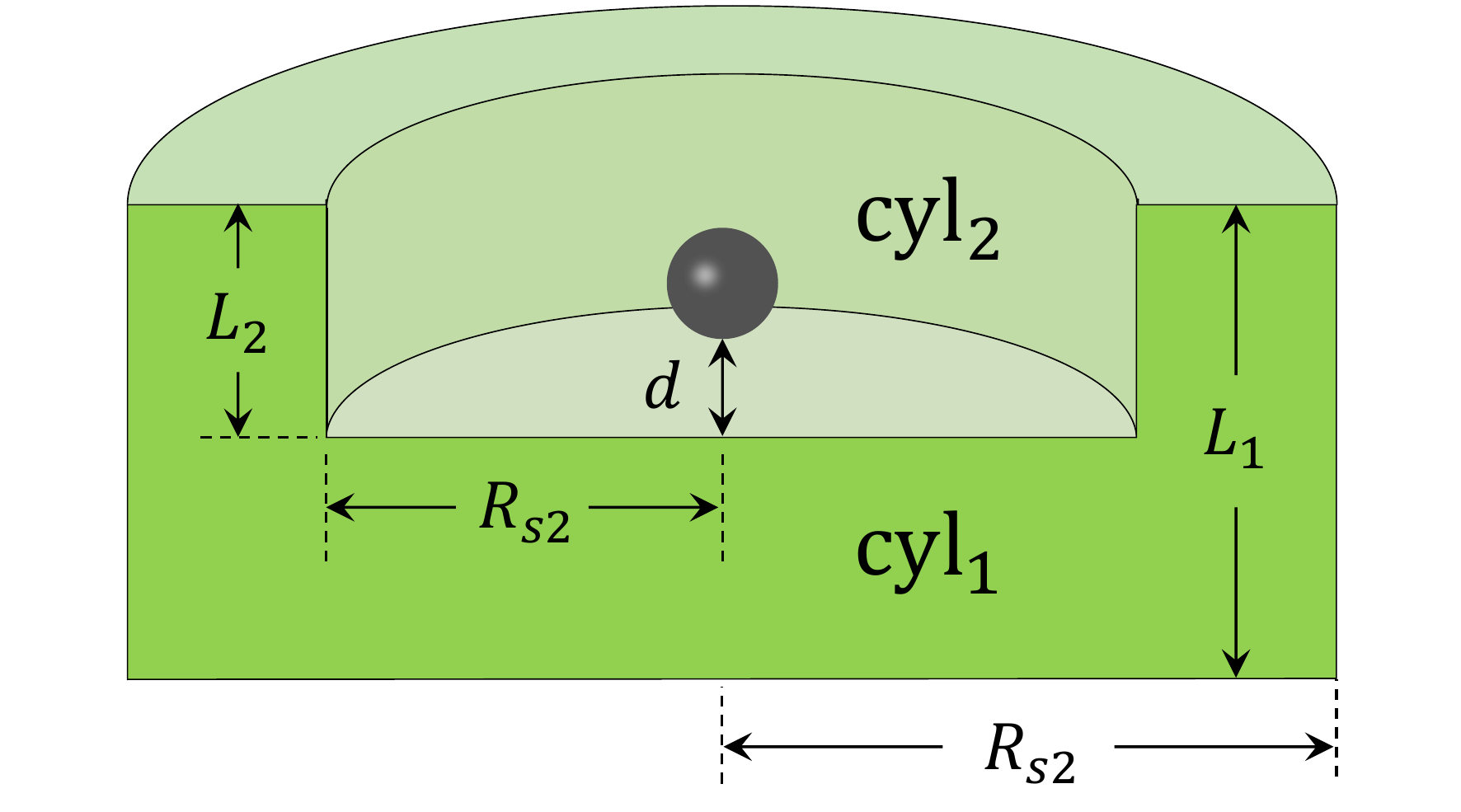}
\caption{The spin source consisting of a large cylinder (cylinder1) with a small cylinder (cylinder2) removed. $R_{\rm s1}$ and $R_{\rm s2}$ are radii of the two cylinders, and $L_1$ and $L_2$ are their heights, respectively. The gray ball represents the microsphere. $d$ is the surface distance between the microsphere and the spin source.}
\label{shape1}
\end{figure}

Apart from the desired magnetic trap, the spin source can induce a magnetic force $F_{\rm s}$ on the microsphere as follows
\begin{align*}
F_{\rm s}=\int_{\rm m}{\rm d}V\frac{\chi_{\rm m}}{\mu_0}\left(B_{\rm 0z}\frac{\partial B_{\rm sz}}{\partial z}+B_{\rm sz}\frac{\partial B_{\rm 0z}}{\partial z}\right).
\end{align*}
This force can be eliminated by deliberately designing the configuration of the spin source (in Fig.~\ref{shape1}).

The $z$ direction component of the magnetic field produced by a single spin is
\begin{align}
B_{\rm sz0}=\frac{\mu_0\mu_B}{4\pi}\frac{3{\cos}^2\theta-1}{l^3},
\label{B_sz0}
\end{align}
where $\theta$ is the polar angle and $l$ is the distance from the microsphere to the spin. The magnetic field of a spin-source cylinder at $z$ axis is then
\begin{align}
B_{sz_i}&=\int_{\rm{cy_i}}{\rm d V}\rho_{\rm{e}}(r_{\rm i},z_{\rm i},t)B_{\rm{sz0}}.
\label{B_szi}
\end{align}
Here $i = 1$ and $2$ correspond to cylinder1 and cylinder 2, $\int_{\rm{cy_i}}{\rm dV}=\int_{-R_{\rm si}}^{R_{\rm si}}{\rm d}z_{\rm i}\int_{0}^{\sqrt{R^2_{\rm si}-z^2_{\rm i}}}2\pi r_{\rm i}{\rm d}r_{\rm i}$, and $\rho_{\rm{e}}(r_{\rm i},z_{\rm i},t)$ is net spin density along the $z$ axis in the cylinder.

The microsphere is assumed to be right above the center of the cylinder, so that the magnetic field in the microsphere is approximately uniform in the transverse direction. Thus, the magnetic force produced by a cylinder on this microsphere is
\begin{align}
F_{\rm s_{{cy}_i}}(t)&=\int_{\rm m}{\rm d V}\frac{\chi_{\rm m}}{\mu_0}\left(B_{\rm 0z}\frac{\partial B_{\rm sz_i}}{\partial z}+B_{\rm sz_i}\frac{\partial B_{\rm 0z}}{\partial z}\right)
\label{F_si},
\end{align}
where $\int_m {\rm d} V$ is the integral over the microsphere. Therefore, the magnetic force produced by the spin source on the microsphere is as follows:
\begin{align}
\notag
F_s(t)&=F_{\rm s_{cy1}}-F_{\rm s_{cy2}}\\
&=\rho_{\rm e}(t)\frac{\chi_{\rm{m}}\mu_{\rm{B}}}{2}\frac{\partial B_{\rm 0z}}{\partial z}\zeta_{\rm s}(d,R),
\label{F_s}
\end{align}
where $\zeta_{\rm s}(d,R)$ is the effective volume for $F_{\rm s}(t)$, reads:

\begin{widetext}
\begin{align}
\notag
\zeta_{\rm s}(d,R)&=\int_{\rm m}{\rm d}V\left(\frac{B_{\rm 0z}}{\frac{\partial B_{\rm 0z}}{\partial z}}\left(\frac{R_{\rm s1}^2}{\sqrt{\left(R_{\rm s1}^2+z'^2\right)^3}}
-\frac{R_{\rm s1}^2}{\sqrt{\left(R_{\rm s1}^2+(z'+L_1)^2\right)^3}}
-\frac{R_{\rm s2}^2}{\sqrt{\left(R_{\rm s2}^2+z'^2\right)^3}}+\frac{R_{\rm s2}^2}{\sqrt{\left(R_{\rm s2}^2+(z'+L_2)^2\right)^3}}\right)\right.\\
&\qquad\left.+\left(\frac{z'}{\sqrt{{R_{\rm s1}}^2+z'^2}}-\frac{z'+L_1}{\sqrt{{R_{\rm s1}}^2+(z'+L_1)^2}}-\frac{z'}{\sqrt{{R_{\rm s2}}^2+(z'+L_2)^2}}+\frac{z'+L_2}{\sqrt{{R_{\rm s2}}^2+(z'+L_2)^2}}\right)\right) .
\label{zeta}
\end{align}
\end{widetext}
In the cylindrical coordinate system, we have $\int_{\rm m}{\rm d}V = \int_{-R}^{R}{\rm d}z\int_{0}^{\sqrt{R^2-z^2}}2\pi r{\rm d}r$, and $z'=z-L_2-d-3R$.

\begin{figure}[htp]
\includegraphics[width=8.6cm]{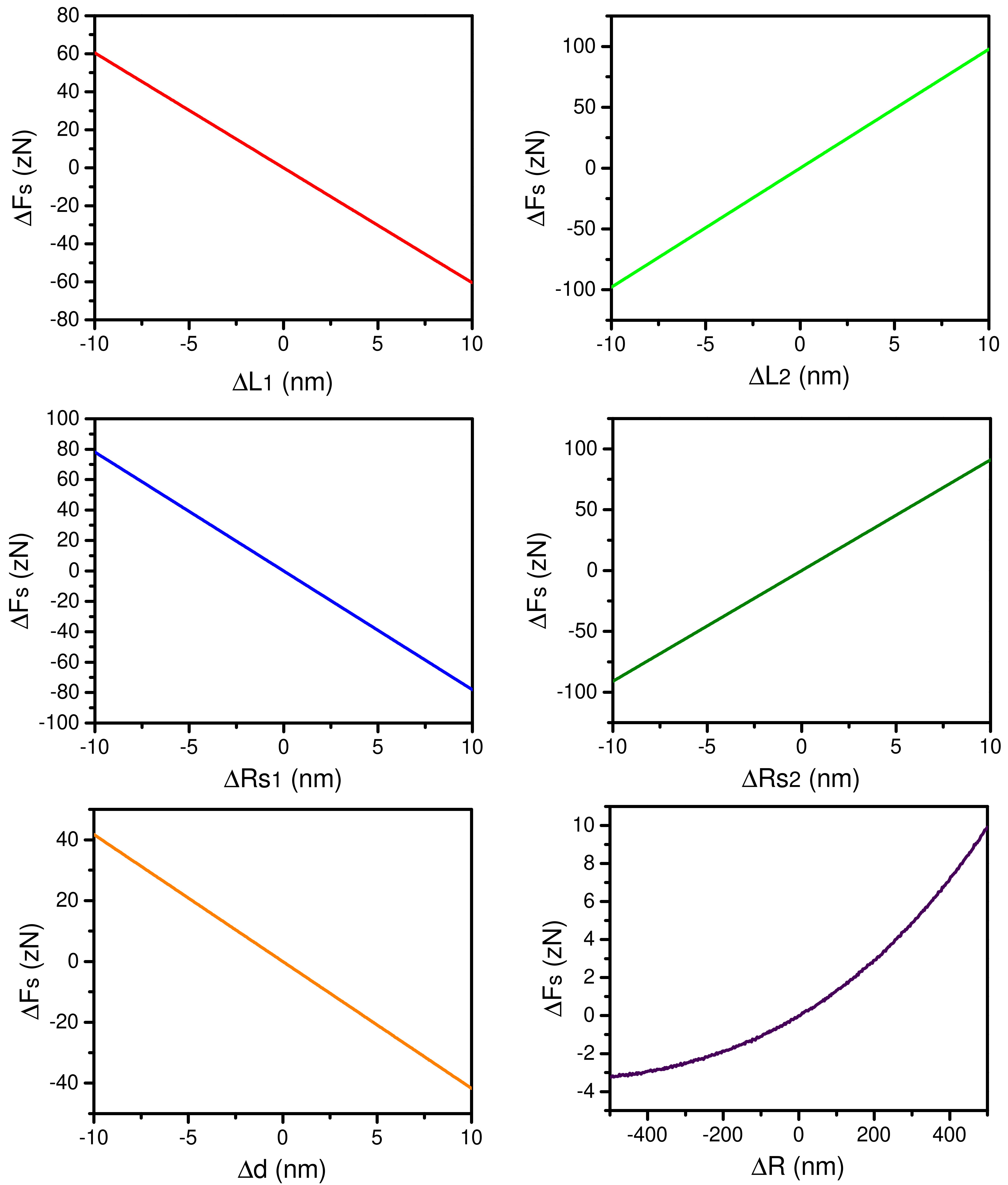}
\caption{Variation of $\Delta F_{\rm s0}$ (standard deviation of $F_{\rm s0}$) with $\Delta L_1,\ \Delta L_2,\ \Delta R_{\rm s1},\ \Delta R_{\rm s2},\ \Delta R,\ \Delta d$ calculated based on Eq.~\eqref{F_s} and Eq.~\eqref{zeta}.}
\label{Error_Range}
\end{figure}

The geometry shape and the imperfections on fabrications are considered. The geometric parameters are optimized to make $F_s$ as small as possible. \autoref{Table1} lists the optimized geometric parameters and their standard deviations according to the practical condition. Here we exaggerate the $\rho_{\rm e}(t)$ to be $\rho_{\rm e0}$. From the table, we can see that the value of optimized $F_{\rm{s}}$ is $4.2\times{10}^{-22}$ N, while the total uncertainty of $\Delta F_{\rm{s}}$  is $\Delta F_{\rm{s}} = 5.03\times{10}^{-20}$ N. More generally, the variation of $\Delta F_{\rm{s}}$ versus the standard deviations of geometric parameters is plotted in Fig.~\ref{Error_Range}.

\begin{table}[htbp]
	\centering
\caption{Structure parameters of the spin source plotted in FIG.\ \ref{shape1} and their effects. The optimized geometrical parameters and their standard deviations are listed in the second column. After optimization, $\zeta_{\rm s} = {\rm {10}^{-23} m^3}$ and $F_{\rm s} = 4.2\times{10}^{-22}$N. $\Delta\zeta_{\rm s}$ and $\Delta F_{\rm s}$ due to parameter uncertainties are also listed. The total uncertainty of $F_{\rm s}$ is listed at bottom right corner, which is far greater than the value of $F_{\rm{s}}$.}
\begin{tabular}{cccc}
        \botrule
  Parameter &\hspace{0.3cm} $\rm{Size(\mu m)}$   &\hspace{0.3cm}$\rm{\zeta_s(10^{-22}m^3)}$ & \hspace{0.3cm}$\rm{F_s(10^{-21}N)}$     \\ \colrule
  $\rm {L_1}$   &\hspace{0.3cm}$59.703\pm0.003$ &\hspace{0.3cm}$-0.1\mp5.0$ &\hspace{0.3cm}$-0.42\mp18.2$      \\
        $\rm{L_2}$   &\hspace{0.3cm}$48.674\pm0.003$ &\hspace{0.3cm}$-0.1\pm8.1$ &\hspace{0.3cm}$-0.42\pm29.3$      \\
        $\rm{R_1}$   &\hspace{0.3cm}$460.00\pm0.003$ &\hspace{0.3cm}$-0.1\mp6.4$ &\hspace{0.3cm}$-0.42\mp23.4$      \\
        $\rm{R_2}$   &\hspace{0.3cm}$440.93\pm0.003$ &\hspace{0.3cm}$-0.1\pm7.5$ &\hspace{0.3cm}$-0.42\pm27.3$      \\
        $\rm{d}$   &\hspace{0.3cm}$1.46\pm0.001$ &\hspace{0.3cm}$-0.1\mp1.1$  &\hspace{0.3cm}$-0.42\mp4.29$      \\
        $\rm{R}$   &\hspace{0.3cm}$3.2\pm0.1$ &\hspace{0.3cm}$-0.1\pm1.3$ &\hspace{0.3cm}$-0.42\pm4.87$      \\\colrule
  $\rm{Total}$   & \hspace{0.3cm}   &\hspace{0.3cm}$\rm{-0.1\pm13.8}$ & \hspace{0.3cm}$\rm{-0.42\pm50.3}$               \\
  \botrule
 \end{tabular}
\label{Table1}
\end{table}

The PSD of the spin-induced magnetic force reads:
\begin{align}
\notag
S_{\rm{ff}}^{\rm{s}}\left(\omega\right)&=\mathscr{F}(\langle F_{\rm s}(t)|F_{\rm s}(0)\rangle)\\
\notag
&=\left(\frac{1}{2}\mu_{\rm{B}}\chi_{\rm{m}}\zeta_{\rm{s}}(R,d)\frac{\partial B_{\rm{0z}}}{\partial z}\right)^2\mathscr{F}(\langle\rho_{\rm e}(t)|\rho_{\rm e}(0)\rangle)\\
\notag
&=\left(\frac{1}{2}\mu_{\rm{B}}\chi_{\rm{m}}\zeta_{\rm{s}}(R,d)\frac{\partial B_{\rm{0z}}}{\partial z}\right)^2\rho^2_{\rm{e_0}}\widetilde{G}\left(\omega\right)\\
\notag
&=(F_{\rm s}+\triangle F_{\rm s})^2\widetilde{G}\left(\omega\right)\\
&\approx (\triangle F_{\rm s})^2\widetilde{G}\left(\omega\right)
\end{align}

\section{PSD OF SPIN-MASS FORCE}\label{sec:C}
The spin-mass effective magnetic field generated by  a polarized spin on an unpolarized nucleon is:
\begin{align}
\bm{B}_{\rm{sp}}(\bm{r})=\frac{\hbar g_{\rm{s}}^{\rm{N}}g_{\rm{p}}^{\rm{e}}}{4\pi m_{\rm{e}}\gamma}\left(\frac{1}{r\lambda}+\frac{1}{r^2}\right)e^{\rm -r/\lambda}\bm{e_{\rm{r}}} .
\label{Bsp}
\end{align}

The spin-mass effective magnetic field generated by the microsphere on a polarized spin is obtained by integrating the volume of the microsphere with Eq.~\eqref{Bsp}, i.e.,
\begin{align}
\bm{B_{\rm{m}}}=\int_{\rm{m}}{\rm d}V\rho_{\rm{m}}\bm{B}_{\rm{sp}}(\bm{r})=\frac{\rho_{\rm{m}}\hbar g_{\rm{s}}^{\rm{N}}g_{\rm{p}}^{\rm{e}}}{4\pi m_{\rm{e}}\gamma}g(R,\ell)\bm{{e}_{\rm{\ell}}},
\label{B_m}
\end{align}
where
\begin{align}
\notag
g(R,\ell)&=2\pi\lambda^2\left(\left(R-\lambda\right)e^\frac{R}{\lambda}\right.\\
&\left.+\left(R+\lambda\right)e^{-\frac{R}{\lambda}}\right)\left(\frac{1}{\lambda\ell}+\frac{1}{\ell^2}\right)e^{-\frac{\ell}{\lambda}} ,
\label{gsm}
\end{align}
$\rho_{\rm{m}}$ is the nucleon density of microsphere, $\bm{e}_{\rm{\ell}}$ and $\ell$ are the unit vector and distance between the microsphere and the spin, respectively. From Eq.~\eqref{B_m} and Eq.~\eqref{gsm} , we can find that in the calculation of spin-mass effective magnetic field, the microsphere is completely equivalent to a center mass. Therefore, Eq.~\eqref{B_m} is equivalent to the effective magnetic field produced by the CM of the microsphere.

The spin-mass potential between the microsphere and the spin-source is obtained by integrating the volume of spin-source with Eq.~\eqref{B_m}:
\begin{align*}
V_{\rm sm}(t)=\int_{\rm{cy_1}}{\rm d}V\rho_{\rm{e}}(r_1,z_1,t)\bm{\mu_{\rm{B}}\cdot B_{\rm{m}}}\\
-\int_{\rm{cy_2}}{\rm d}V\rho_{\rm{e}}(r_2,z_2,t)\bm{\mu_{\rm{B}}\cdot B_{\rm{m}}},
\end{align*}
and $\rho_{\rm{e}}(r_{\rm i}, z_{\rm i}, t)$ represents the net electron spin density along the $z$ axis in the spin-source.

Consequently, the spin-mass force between the microsphere and the spin-source is
\begin{align}
\notag
F_{\rm sm}(t)&=-\frac{\partial V_{\rm sm}}{\partial z}\\
\notag
&=-\frac{\partial}{\partial z}\left(\int_{\rm{cy_1}}{\rm d}V\rho_{\rm{e}}(r_1,z_1,t)\bm{\mu_{\rm{B}}\cdot B_{\rm{m}}}\right.\\
\notag
&\left.-\int_{\rm{cy_2}}{\rm d}V\rho_{\rm{e}}(r_2,z_2,t)\bm{\mu_{\rm{B}}\cdot B_{\rm{m}}}\right) ,
\label{Fsm}
\end{align}
where $\zeta_{\rm{sm}}(R,d,\lambda)$ is the effective volume for $F_{\rm sm}(t)$, reads:
\begin{align}
\zeta_{\rm{sm}}(R,d,\lambda)=(2\pi \lambda)^{\rm{2}}\left((R-\lambda)e^{-\frac{d}{\lambda}}+(R+\lambda)e^{\rm{-\frac{2R+d}{\lambda}}}\right) .
\end{align}

Accordingly, The PSD of spin-mass force reads:
\begin{align}
\notag
S_{\rm{ff}}^{\rm{sm}}\left(\omega\right)&=\mathscr{F}(\langle F_{\rm s}(t)|F_{\rm s}(0)\rangle)\\
\notag
&=\left(\frac{\hbar^2g_{\rm{s}}^{\rm{N}}g_{\rm{p}}^{\rm{e}}\rho_{\rm{m}}}{8\pi m_{\rm{m}}}\zeta_{\rm sm}(R,d,\lambda)\right)^2\mathscr{F}(\langle\rho_{\rm e}(t)|\rho_{\rm e}(0)\rangle)\\
&=\left(\frac{\hbar^2g_{\rm{s}}^{\rm{N}}g_{\rm{p}}^{\rm{e}}\rho_{\rm{m}}}{8\pi m_{\rm{m}}}\zeta_{\rm sm}(R,d,\lambda)\right)^2\rho^2_{\rm{e_0}}\widetilde{G}\left(\omega\right) .
\end{align}

\section{CALCULATION OF $(g_{\rm{s}}^{\rm{N}}g_{\rm{p}}^{\rm{e}})_{\rm limit}$}\label{sec:E}

To observe the spin-mass signal, $g_{\rm{s}}^{\rm{N}}g_{\rm{p}}^{\rm{e}}$ needs to be no less than
\begin{align*}
\notag
(g_{\rm{s}}^{\rm{N}}g_{\rm{p}}^{\rm{e}})_{\rm limit}&=\sqrt{\frac{S_{\rm ff}^{\rm s}(\omega_{\rm z})}{\widetilde{G}(\omega_{\rm z})}}\frac{8\pi m_{\rm e}}{\zeta_{\rm sm}\hbar^2\rho_{\rm m}\rho_{\rm e_0}}\\
&=\frac{\frac{\chi_{\rm{m}}\mu_{\rm{B}}}{2}\frac{\partial B_{0z}}{\partial z}\zeta_{\rm{s}}(d,R)}{\frac{\hbar^{\rm{2}}\rho_{\rm{m}}}{8\pi m_{e}}\zeta_{\rm{sm}}(R,d,\gamma)}.
\end{align*}

For the worst situation, $(g_{\rm{s}}^{\rm{N}}g_{\rm{p}}^{\rm{e}})_{\rm limit}$ takes its upper bound:
\begin{align}
\rm{sup}\left((g_{\rm{s}}^{\rm{N}}g_{\rm{p}}^{\rm{e}})_{\rm limit}\right)=\frac{\rm{sup}\left(\frac{\chi_{\rm{m}}\mu_{\rm{B}}}{2}\frac{\partial B_{\rm{0z}}}{\partial z}\zeta_{\rm{s}}(d,R)\right)}{\rm{min}\left(\frac{\hbar^{\rm{2}}\rho_{\rm{m}}}{8\pi m_{\rm{e}}}\zeta_{\rm{sm}}(R,d,\gamma)\right)},
\label{worst}
\end{align}
where $\rm{sup}\left(\frac{\chi_{\rm{m}}\mu_{\rm{B}}}{2}\frac{\partial B_{\rm{0z}}}{\partial z}\zeta_{\rm{s}}(d,R)\right)$ means the upper bound of $\frac{\chi_{\rm{m}}\mu_{\rm{B}}}{2}\frac{\partial B_{\rm{0z}}}{\partial z}\zeta_{\rm{s}}(d,R)$, and $\rm{min}\left(\frac{\hbar^{\rm{2}}\rho_{\rm{m}}}{8\pi m_{\rm{e}}}\zeta_{\rm{sm}}(R,d,\lambda)\right)$ is the minimum value of $\frac{\hbar^{\rm{2}}\rho_{\rm{m}}}{8\pi m_{\rm{e}}}\zeta_{\rm{sm}}(R,d,\lambda)$. We take
\begin{align}
&\rm{sup}\left(\frac{\chi_{\rm{m}}\mu_{\rm{B}}}{2}\frac{\partial B_{\rm{0z}}}{\partial z}\zeta_{\rm{s}}(d,R)\right)\notag\\
=&\frac{\chi_{\rm{m}}\mu_{\rm{B}}}{2}\frac{\partial B_{\rm{0z}}}{\partial z}\left(\zeta_{\rm{s}}(d,R)\right.
\left.+\triangle \zeta_{\rm{s}}(d,R)\right) \notag\\
\approx & \triangle F_{\rm s} .
\label{deltaF}
\end{align}
and $\rm{min}\left(\frac{\hbar^{\rm{2}}\rho_{\rm{m}}}{8\pi m_{\rm{e}}}\zeta_{\rm{sm}}(R,d,\lambda)\right)$ is numerically calculated with parameters $R$ and $d$ taken within the uncertainty ranges (see \autoref{Table1}). Combined with Eq.~\eqref{worst} and Eq.~\eqref{deltaF}, the estimated  $(g_{\rm{s}}^{\rm{N}}g_{\rm{p}}^{\rm{e}})_{\rm limit}$ in the worst situation is shown in red in Fig.~\ref{fig3} in the main text.


%

\end{document}